\documentclass[quaderni]{sifrev}        
\usepackage{geometry}
\usepackage{amsmath}
\usepackage{footnote}
\usepackage{footmisc}
	\renewcommand\and{\mbox{\rm and\ }\ignorespaces}

\begin{document}

\title{Enrico Fermi in Argentina and his\\Lectures in Buenos Aires, C\'ordoba and La Plata}

\shorttitle{Enrico Fermi in Argentina...}

\author{Alessandro De Angelis\thanks{E-mail: alessandro.deangelis@unipd.it}\inst{1} \and Jos\'e M. Kenny\inst{2}}           

\institute{Dipartimento di Fisica e Astronomia ``Galileo Galilei'' dell'Universit\`a di Padova\\
Istituto Nazionale di Fisica Nucleare \& INAF, Padova, Italy\\
Universidade de Lisboa \& Instituto Superior T\'ecnico/LIP,  Lisboa, Portugal \and Ambasciata d'Italia in Argentina,  Buenos Aires\\
Universit\`a di Perugia, Italy}


\vol{42}                                  
\issue{2}				  

\month{Month year}                        


\riassunto{Nel 1934 Enrico Fermi accett\`o un invito a tenere conferenze in Argentina, Brasile e Uruguay. Arrivato  il 30 luglio,
parl\`o a Buenos Aires, C\'ordoba, La Plata e Montevideo, e poi si spost\`o il 18 agosto a San Paolo passando per Santos e Rio de Janeiro; ritorn\`o da Rio a Napoli il primo settembre. La sua visita ebbe una grande risonanza e le aule erano affollate nonostante  tenesse le sue lezioni in italiano. 
 L'Universit\`a di Buenos Aires registr\`o le sue cinque lezioni e le  trascrisse in spagnolo.
 Contengono le prime presentazioni pubbliche della teoria del decadimento $\beta$ e dei lavori sulla radioattivit\`a artificiale del gruppo via Panisperna, ma
    non sono incluse fra i {\em Collected Works} di Fermi a cura dell'Accademia dei Lincei di Roma e dell'Universit\`a di Chicago,
 sebbene elencate nella bibliografia. 
 In questo articolo presentiamo la trascrizione delle  lezioni di Fermi a Buenos Aires, un riassunto della lezione a La Plata e
 un ampio riassunto della lezione a C\'ordoba, traducendo questo materiale per la prima volta in inglese.

}

\abstract{{In 1934 Enrico Fermi accepted an invitation to  lecture in Argentina, Brazil and Uruguay. He arrived in Buenos Aires on July 30,
 lectured in Buenos Aires, C\'ordoba, La Plata and Montevideo,  and then moved  on August 18 to S\~ao Paulo in Brazil via Santos and Rio de Janeiro; he traveled back from Rio to Naples on September 1st.  His visit had a large resonance, and halls    were crowded despite the fact that he lectured in Italian. The University of Buenos Aires recorded his five lectures and transcribed them in Spanish.
 They contain the first public presentations of the theory of $\beta$ decay and of the works on artificial radioactivity started by the via Panisperna group, but
    are not included in Fermi's {\em Collected Works} edited by the Accademia dei Lincei in Rome and by the University of Chicago,
 although listed in the Bibliography. In this paper we present the transcription of Fermi's   lectures in Buenos Aires, a summary of the lecture in La Plata and 
 an extended summary of the lecture in C\'ordoba, translating them in English for the first time. 
}}

\maketitle

\newpage


\newpage


\addcontentsline{toc}{section}{Enrico Fermi in Argentina}

Enrico Fermi (Rome 1901 - Chicago 1954) has been one of the most important scientists ever, both for his contributions to theoretical and experimental physics. Graduated in 1922 at the University of Pisa and awarded the same year the Diploma of the Scuola Normale Superiore  discussing a thesis on X-rays, in 1925 he was appointed adjunct professor of mathematical physics in Florence. In 1926,  he invented the so-called Fermi-Dirac statistics that describes the energy-level distributions of semi-integer spin particles, such as electrons, protons, neutrons and other particles today generically called fermions. Thanks to this work, and to his extraordinary overall scientific production, he was assigned in 1926 the chair of theoretical physics in Rome, the first in Italy. In the autumn of 1926 he moved to Rome in the Institute of via Panisperna, where he began the most fruitful period of his scientific life, and soon created a group of collaborators (Rasetti, Segr\`e, Amaldi, Majorana, Pontecorvo, D'Agostino). The   via Panisperna group initially dealt with atomic spectroscopy and later with the study of the atomic nucleus.  After the discovery of the neutron by Chadwick in 1932, between 1933 and 1934 Fermi elaborated the $\beta$-decay theory, based on the physical hypothesis advanced by Pauli in 1930 that  
a neutral, light and 
invisible particle escaping detection is present in  the final state of the decay; this particle is called today
 (anti)neutrino (the term ``neutrino'' was invented by Fermi). His theory highlighted for the first time the existence of a new force, today called weak interaction. In 1934, hearing of the discovery by Curie and Joliot of radioactivity caused by bombarding stable nuclei with $\alpha$ particles ($^4$He nuclei), he began a series of experiments on the disintegration of nuclei using neutrons produced in radon-beryllium sources. Neutrons, being neutral, proved more effective than the $\alpha$ particles for penetrating nuclei and modifying or breaking them, and Fermi soon discovered nearly forty new artificial radioactive nuclides. In particular, he discovered the great effectiveness of slow neutrons (obtained by  fast neutrons colliding with light nuclei) in producing  $\beta$-radioactive transitions. With slow neutrons, by the end of 1934 the new radioactive nuclides found by Fermi were over seventy. In 1938 he went to Stockholm to receive the Nobel Prize, awarded for his fundamental works on neutrons, and from there he moved on to the US where he settled. In 1939 he became a professor in Chicago; in the same year he demonstrated that the secondary neutrons produced in the nuclear fission, discovered by Hahn, Meitner and Strassman in 1938, could in turn produce new fissions and give rise to a chain reaction with the release of nuclear energy at  a macroscopic level, and he designed the first nuclear reactor, which entered into operation in  December 1942. He was one of the protagonists of the Manhattan project for the  use of nuclear energy in the atomic bomb, and in 1944 he became an American citizen. After the war 
 Fermi devoted himself to the theory of cosmic ray 
acceleration (reaching a qualitatively correct formulation), to 
experiments with the renowned Chicago cyclotron where he discovered the 
first hadronic resonance, the $\Delta^{++},$ and to the development of computers \cite{treccani}.

Year 1934 was the {\em annus mirabilis} for  Fermi \cite{bernardini}, with many great discoveries, in particular the theory of $\beta$ decay (the result had been indeed made public
already in 1933 \cite{betars}, but the complete publication came in 1934 \cite{betanc}), which paved the way to the theory of electroweak interactions and to the Standard Model of particle physics, and neutron-induced artificial radioactivity, that will lead him to the Nobel prize in 1938
(for his demonstrations of the existence of new radioactive elements produced by neutron irradiation, and for his related discovery of nuclear reactions brought about by slow neutrons). In the same year  Fermi  accepted an invitation to deliver lectures in Argentina, Brazil, and Uruguay. He reached Buenos Aires on Monday, July 30 1934   onboard the Neptunia\footnote{An interesting coincidence: Fermi had 
possibly discovered just before his travel to South America the new element with atomic number 93, which is called, today, neptunium (Np), but he  had  doubts. He discussed 
about hints of this discovery during his 5th lecture in Buenos Aires, and in C\'ordoba.
The discovery of neptunium by Fermi was  controversial. 
Fermi's team bombarded uranium with neutrons and observed evidence of a  resulting isotope with atomic number  93. In 1934 he proposed the name ausonium (atomic symbol Ao) for element 93, after the Greek name Ausonia (one of the Greek  names of Italy). Several  objections to his claims were raised; in particular, the exact process that took place when an atom captured a neutron was not well understood. Fermi's  accidental discovery   cast further doubt in the minds of many scientists: he  could not prove that element 93 was being created unless he could isolate it chemically, and he could not. Otto Hahn and Lise Meitner,  among the best radiochemists in the world at the time and supporters of Fermi's claim, unsuccessfully tried. Much later, it was determined that the main reason for this failure was due to the fact that the predictions for element 93's chemical properties were based on a periodic table  lacking the actinide series. While the question of whether Fermi's experiment had produced element 93  had no clear answer yet (Fermi had likely and unknowingly observed nuclear fission reactions), in early 1939 Edwin McMillan at the Berkeley Radiation Laboratory of the University of California, Berkeley, decided to run an experiment bombarding uranium using the  cyclotron that had recently been built at the university, producing beyond any doubt the  new element.}  ship of the Cosulich company leaving from Trieste \cite{ministero}, but stopping also in Naples, where Fermi probably boarded. The most important Italian newspaper, {\em Corriere della Sera}, wrote about the visit on its cultural page (page 3) on  July 24 \cite{corriere}.

According to the memories of his wife \cite{laura}  Laura, Fermi {\em could not break his engagement at the last moment to pursue experiments, no matter how absorbing and promising.} Anyway {\em it would have been a mistake to forego this trip, which proved most successful from all points of view. Sixteen days on placid seas} took the Fermis to Buenos Aires. 

\begin{figure}
\begin{center}
\includegraphics[width=.3\textwidth]{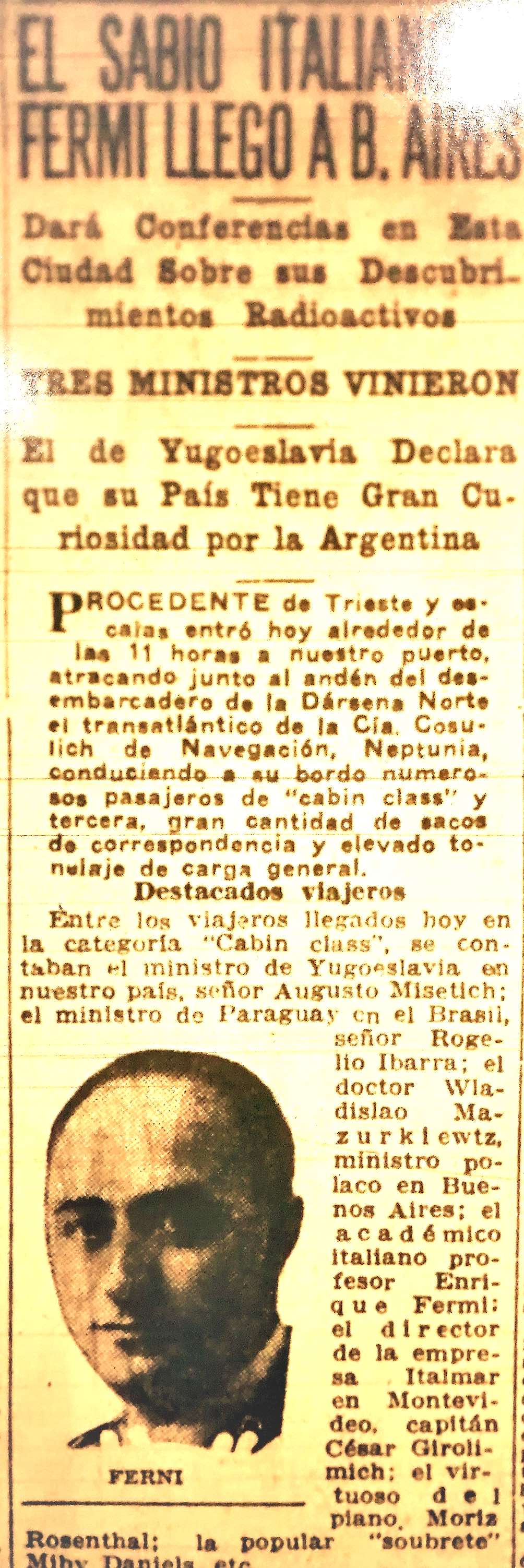}
\end{center}
\caption{\label{fig:giornale} Article on the first page of the daily newspaper ``Cr\'itica'' dated July 30, 1934 \cite{critica}.}
\end{figure}

Fermi was well awaited and received in Buenos Aires. The local press gave large resonance to the arrival of the scientist  (Figure \ref{fig:giornale}, \cite{critica}, \cite{mattino}, ...), and the Fermi couple experienced a sumptuous hospitality.
As Laura  wrote \cite{laura}, {\em  we lived the life of the elite for over three weeks. We were housed in the most modern and elegant hotel we had seen up to then. Introduced by the Italian ambassador and by the president of the Instituto Argentino de Cultura Italica, the sponsor of Enrico's visit, we were entertained in the upper spheres of New World success and wealth. Perhaps out of true interest in science or because of a vague nostalgia for the culture of the old world that they could not forget, many prominent citizens of Buenos Aires seemed eager to shed their kindness on us. They took us for rides along the Rio de la Plata, and up to the Paran\'a; they invited us into their theater boxes for the best shows and musical performances; they entertained us in their sumptuous homes with that proverbial Spanish hospitality, so hard on the guests' digestive system, that makes a hostess place another guest in charge of keeping your plate full at a meal of five courses -- we came to look forward to the rare occasions when, free of invitations, we could quietly skip a meal. A great testimony of interest in science was given to Enrico.}
Fermi held his lectures in halls that were crowded and overflowing at the onset and kept overflowing to the very end of his course, despite the fact that he lectured in Italian \cite{laura,segre}. {\em [A] good portion of the Buenos Aires population is of Italian descent.\footnote{During the time in which Fermi stayed in Buenos Aires, the tenor Tito Schipa performed in Teatro Colon and the director Ottorino Respighi in Teatro Cervantes.}} 

Fermi gave five lectures in Buenos Aires, starting on Thursday, August 2. The  subjects of the lectures were: 
\begin{enumerate}
\item[I.] Characteristics distinguishing atomic physics from the physics of ordinary bodies. 
\item[II.] The concept of measurement and its criticism.
\item[III.] The fundamental elements of nuclear structure (part 1).
\item[IV.] The fundamental elements of nuclear structure (part 2).
\item[V.] The artificial disintegration of the nucleus.
\end{enumerate}

The lectures were recorded by the University of Buenos Aires \cite{ba}, but have  not been included in the {\em Collected Works}  of Enrico Fermi \cite{opere}, although listed in the bibliography (table of contents) with a star. The Foreword of the  {\em Collected Works} says:  ``The papers marked with a star have not been reprinted, either because they are translations or abbreviations of papers which are reprinted, or because they are hastily composed accounts, often written by other persons, of Fermi popular lectures." The Buenos Aires lectures belong to this second category, but the account is certainly not hasty, although (and this is unusual) written in Spanish. They are extremely important, in particular because they include the first transcribed public presentation of the theory of $\beta$ decay and of the works on artificial radioactivity started by the via Panisperna group.\footnote{Neutron-induced radioactivity had already been presented once by Fermi in a public lecture in May 1934 at the Institute of Radiology in Rome. This was however not transcribed.} We present in the appendix the first English translation.

\begin{figure}[h]
\begin{center}
\includegraphics[width=.4\textwidth]{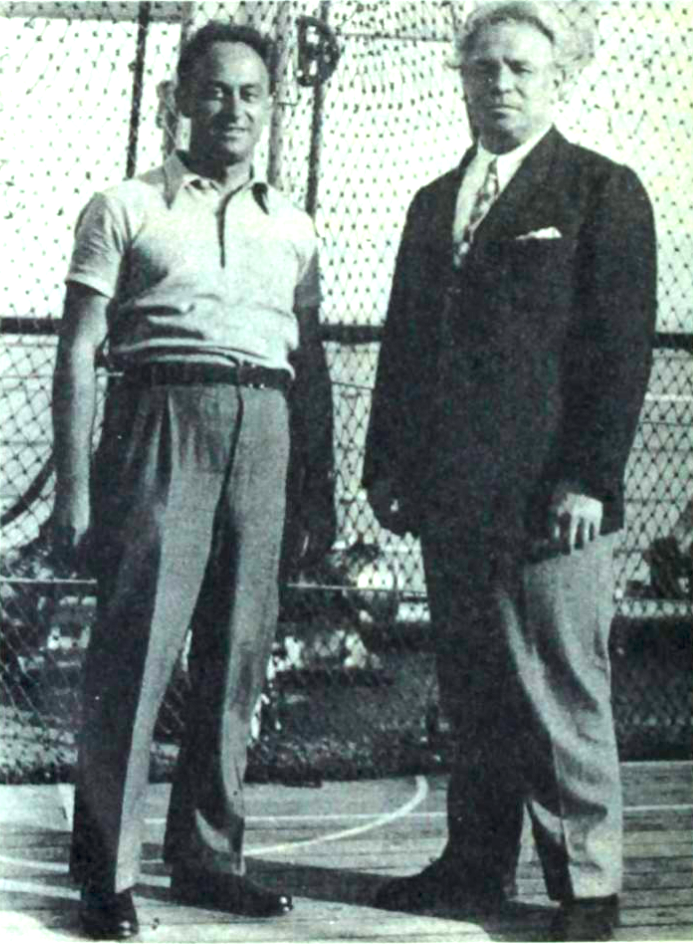}
\end{center}
\caption{\label{fig:respighi}  Fermi and Ottorino Respighi in the ship coming back from South America. From \cite{laura}.}
\end{figure}

On  Saturday, August 11, Fermi gave in C\'ordoba a lecture on artificial radioactivity (an extended summary in Spanish is available in \cite{cordoba}; we translated also this  in the appendix to the present work), 
and later  in Montevideo. Also  {\em in small C\'ordoba, the town of the many churches at the feet of the Andes, where the only Italian was a fencing teacher,} [and] {\em in orderly, green-gardened, intellectual Montevideo,} lecture rooms were tightly packed \cite{laura}. 

A last Argentinean lecture on Radiation (radioactivity, light  and cosmic rays), held on Friday, August 17, in the
University of La Plata, is documented; a summary
is available \cite{laplata} and is translated in the appendix. Also in La Plata many students, professors and ordinary people attended. 

On August 18 Fermi left Argentina to Santos with the ship Northern Prince arriving on the 21st, and then to S\~ao Paulo,  {\em  in whose surroundings the intense green of the tropical vegetation sprang from the bright-red soil that gives Brazil its name in an antithesis of colors seldom achieved by painters} \cite{laura}, and Rio de Janeiro. The time spent
in Brazil, invited by the mathematician Theodoro Augusto Ramos, professor in the Escola Politecnica of S\~ao Paulo, is documented in \cite{brazilians,brazilians2}. Brazil was looking for help in starting a school in theoretical physics, and Fermi was functional to the call of Gleb Wataghin from Torino for a chair in theoretical physics in S\~ao Paulo \cite{brazilians}. We note that Fermi's lectures in Brazil have not been transcribed.

Finally the staying in South America came to an end, and the Fermi family left  Rio to Naples on September 1st. We cite again the words of Laura \cite{laura}. {\em We left South America happy to end an experience too rich to be stretched any further. Of good things one must not have too much, and, as the French say, one may tire of eating partridge every day. On boarding the boat that would take us from Rio to Naples, we met the composer Ottorino Respighi and his wife.} Fermi and Respighi (Figure \ref{fig:respighi}),
companions for all the trip,   discussed most of the time, with Fermi trying to extract information on a possible mathematical theory of music, and Respighi, according to Laura, smiling with condescension. The boat  arrived in Naples during the second half  of September; then Fermi's wife went to rest in Florence at her aunt's villa while Enrico went to Rome alone, where some of his greatest discoveries expected him, notably radioactivity induced by  neutrons, for which he will be awarded the Nobel prize in 1938.


\subsection*{Acknowledgements}
Important information on the visit by Fermi in South America was provided by Ugo Amaldi,  Gianni Battimelli, Luisa Bonolis, Adele La Rana.  Ron Shellard and C\'assio Leite Viera provided us with information on the visit to Brazil. The authors are grateful to the Dean of the Faculty of Exact Sciences of the University of Buenos Aires, Juan Carlos Reboreda, who provided a copy of the transcriptions of Fermi Lectures edited in Spanish by the same Faculty and conserved in its library. Angela Oleandri, Andrea Vitturi and Silvia Lenzi checked and improved the manuscript. A special acknowledgement goes to Luisa Cifarelli, who made very detailed comments and criticisms, injecting in this paper a lot of quality.

\clearpage
\newpage


\setcounter{footnote}{0}
\renewcommand{\thefootnote}{\alph{footnote}}

\section*{{\underline{Appendix}}}

\vskip 1.5 cm

\section*{Buenos Aires: Lecture I\\Characteristics distinguishing atomic physics from the physics of ordinary bodies (August 2, 1934)}

\vskip 0.3 cm

\addcontentsline{toc}{section}{Enrico Fermi's lectures in Buenos Aires (August 2 --)}

\addcontentsline{toc}{subsection}{I: Characteristics distinguishing atomic physics from the physics of ordinary bodies}

First, I would like to express my deepest gratitude to the Argentinean Institute of Italian Culture and to its President, Engineer Marotta, for the high honor and the great pleasure they have given me by inviting me to come to Buenos Aires. I would also like to thank the Faculty of Exact Sciences and its Dean Engineer Butty, as well as Engineer Selva, and Professor Isnardi, for the kind words they said in my regard. It is a real pleasure for me to get to know this country, where so many of my compatriots have found a new homeland, as well as to have the opportunity to meet personally the Argentinean colleagues of whom I only knew the works through publications.

In this lecture, I will deal with the development of our knowledge about the theory of the structure of matter. The problem of the structure of matter has been proposed since the origins of scientific thought, although its true development has been carried out mainly at the end of  last century and has made a breakthrough in our century that could be called explosive. Physical science is probably as old as human thought and probably arose due to two different motivations: a rather practical one, which puts itself at the service of concrete and immediate ends, and another, more speculative, of satisfying curiosity and the quest for knowledge. These two tendencies are mixed from the beginning in physics, as well as in other sciences. Geometry, for example, is in itself one of the most abstract parts of mathematics, but its name indicates that it originated from a practical problem: the measurement of land.

Since ancient times we have evidence of sciences which have been motivated by scientific curiosity, astronomy being the most characteristic example; while in the field of physics especially mechanics, which is a practical science, developed in ancient times. But next to mechanics we find the first attempts at speculations not due to the desire to obtain concrete results. It is sufficient to recall the investigations of the school of Democritus of Abdera, who, for the first time, spoke of the ``atom'' when building his atomic theory. He came to envision this concept guided by the observation of the compressibility of bodies. His way of reasoning was as follows: being unable to conceive that matter without cavities could vary in volume, and seeing that most bodies could vary in volume, he deduced that there were interstices in their structure and that matter was constituted by particles (atoms) separated by gaps. When matter is compressed, the various atoms that constitute it get closer; when it expands, they separate. Naturally, it took many centuries of thought and scientific work before the conception of Democritus could reach more important conclusions; this idea should be considered as a brilliant intuition for his time, if you keep in mind that it was reached with a truly scarce experimental base.

It was after  Middle Ages, in the Modern Age, that the real development of science began. Here again we can see that the first sciences that developed were those concerning the phenomena that most directly impress our senses: mechanics and optics, and later, the science of heat. The last to develop was electricity, probably because electrical phenomena were the least evident. It is known, for example, that in the open air there is a fairly considerable electric field, which however entirely escapes  our senses and although there are other electrical phenomena that are very apparent, such as lightning, they obviously do not adapt to being taken as starting point for a science. All this explains why the study of electricity has developed late, in the last century.

In the study of electricity, we also find the combination of several trends, some speculative, other experimental or technical, which culminated finally in a set of concrete and useful results for the needs of mankind. The most prominent example is radio communications that made such brilliant technical progress in our century. This is also a case of great practical importance, originated from a purely speculative field: Maxwell, summarizing the results of electromagnetism, established equations that he later managed with purely mathematical means and found that there is the possibility that electromagnetic waves propagate. In his speculations he did not go beyond the ultra-theoretical field. The second stage, when moving on to the experimental phase, is due to Hertz, who took up Maxwell's works, experimentally demonstrating the existence of the waves theoretically foreseen by this physicist. The last stage is that of Marconi, more practical-minded  than the other two. Marconi wondered: what can these waves do? Under the conditions established by Hertz, waves could not travel a distance greater than a few meters; Marconi tried to go further and take advantage of them to transmit signals. Theorists said to Marconi: consider that the Earth is curved, and therefore you can only telegraph as far as the waves can travel straight, and not beyond. Marconi was not a theoretician, and these objections did not worry him excessively. He moved on and made transmissions up to such distance where they could be picked up, so far away that theorists affirmed,
holding their head in their hands, that it was not possible. A circumstance had been ignored: the existence of a conductive electrical shell in the atmosphere, which encloses the atmospheric layer in which we live, between two surfaces: the surface of the Earth and that of that outer shell separated from it by a very small distance compared to the radius of the Earth. When waves are transmitted in the atmospheric layer, they are forced within this layer and  can reach the antipodes. We have therefore started from mathematics and then, through physics, we came to the technique.

By encouraging science, even the most abstract one, tomorrow's technique is being prepared. Of course the promotion of science is a long-term capital investment, since experience shows that the technique follows a long way from the scientific results; but in the life of Nations, investments can be made over very long times.

Let us now return to the origin of atomic theory. If the argument for the existence of atoms were only Democritus' compressibility, this hypothesis would certainly seem very dubious. On the other hand, modern science has collected today such a significant evidence in its favor that having doubts about it is no longer rationally possible. This evidence has been found only in modern times, when chemistry ceased to be alchemy and became a quantitative science, freeing itself from the prejudices that obscured its birth. Chemistry thus managed to establish the laws of multiple and defined proportions. These laws are interpreted in the simplest way, admitting that there are some fundamental substances, the chemical elements, and that each of these substances is made up of very small units, so small that they can only be seen in large quantities. Then, the molecules of the innumerable compounds  are formed by taking determined numbers of these fundamental units of each element.

In this atomic chemical approach, we still miss a way to measure the absolute size of the molecules. Molecules and atoms are not visible: they are of submicroscopic dimensions, surely smaller than one tenth of a thousandth of a millimeter. If only chemistry existed as the basis of the atomic approach, a criterion would be lacking to specify how much smaller the atoms actually are than the above limit, but the atomic concept is also reached by another route, which is precisely the study of the properties of gases.

The molecular kinetic theory of gases gives a truly simple and convincing explanation of the properties of gases themselves, and in particular of their tendency to expand. If the gas is made up of a large number of continuously moving corpuscles, they will invade all the regions that are free and available. In this kinetic theory of gases we find for the first time an argument that allows us to determine the molecular dimensions. The internal friction of the gas depends on the molecular dimensions, in such a way that the measurement of either quantity implies the measurement of the other. The dimensions found with this and other equivalent methods are very small. To get an idea, you can imagine a bullet of diameter 2/3 of a centimeter compared to the diameter of the Earth: this would be the size of an atom compared to the length of one meter.\footnote{This  is probably  transcribed  badly in the Spanish text.  To let the audience figure out how small an atom was, Fermi,  famous for his back-of-the-envelope calculations, compared a bullet of $6.7 \times 10^{-3}$ m diameter to the size of the Earth, whose diameter is about $1.3 \times 10^7$ m. The ratio $D_{\rm bullet}/D_{\rm Earth}$ is about $5 \times 10^{-10}$, a number which corresponds in meters to a typical atomic diameter. The text in Spanish says ``Para darse una idea se puede imaginar a una munici\'on de 2/3 centimetro de di\'ametro aumentada hasta las dimensiones de la Tierra, conservando inalterada su estructura, en estas condiciones un \'atomo resultaria del tama\~no de la munici\'on'' (note of the editors).}

Sometimes we are forced, intuitively, to consider as real only what we can see or feel directly. Given the smallness of their dimensions, we can never hope to see molecules. In fact, although you can think about perfecting the microscope technique, you will never be able to clearly distinguish objects with dimensions less than the wavelength of light, which, although small, is thousands of times larger than the atom. Nevertheless, if you can't see atoms with ordinary light, you can study them with X-rays that are nothing but light of a different color. They allow, if not to see in the true sense of the word, at least to do something similar.

With the experiment of Laue and the Braggs on the structure of crystals, it was indeed possible to accurately recognize the position of atoms and molecules in a crystalline body as if the objects were viewed under a microscope. 

The image of a gas that results from the kinetic theory is something very disordered: the molecules collide irregularly with each other and against the walls, describing very complicated motions, so it may seem strange that the gas as a whole obeys simple and very precise laws. The reason for this simplicity must be sought in the fact that the molecules are in very large numbers, so that in all the phenomena in which the properties of a set are studied, the individual irregularities disappear on the average. This regularization of the laws relating to a very large group is a completely general phenomenon of mathematical nature. Just as we find regular laws in the statistics of molecules, we can find them in the statistics of a social or economic phenomenon regarding the properties of a large population. From parents to children we find fluctuations in height of a completely irregular nature, which seem to escape all laws, but if we take the average height of a people we find that individual fluctuations disappear. We can find variations in height, but the phenomenon will be justifiable with the variation of economic conditions or variations in climate: the average destroys all the capricious individual variations.

The quantitative study of the properties of a gas through the analysis of the behavior of each molecule is a mathematically insoluble problem. However, statistical mechanics forgoes the search for such a solution and accepts as a valid alternative the average properties of a large number of molecules. These average properties determine the behavior of a gas as we observe it. Sometimes the irregular movements 
produced by the irregularity of molecular motion  are observed under the microscope. If very small colloidal particles are observed microscopically, it will be seen that they are in continuous motion. If we observe a large body in the calm air, we will see that it remains still, because it receives shocks on its entire surface that push it in all directions, canceling their effects on average, so that the body remains immobile. However, if the body is small, the number of  shocks is not so great and therefore one can perceive slight differences in the intensity of the shocks in one direction or another, which produce a visible motion of these corpuscles. These movements take the name of Brownian movements.

Knowing the existence of atoms, the question naturally arises  of whether these corpuscles should be considered truly indivisible, as in the classical understanding, or if they possess a complex structure. An issue of this kind could not be discussed if not on the basis of experiments, and the first experiments that could have made us suspect the existence of an internal structure of the atom were performed in the second half of the last century. The phenomena of electrolysis, on the one hand, demonstrate that, in certain circumstances, atoms can give rise to ions electrically charged, a fact that indicates the presence of charged corpuscles inside. The existence of these subatomic charged corpuscles is more directly evidenced by the study of electrical discharge in rarefied gases, and particularly in the study of cathode rays. These turned out to be an emission of particles of extremely small mass and of negative electrical charge, which were called electrons and are present as common constituents in all atoms. Indeed the same electrons with the same negative electrical charge and with the same mass are obtained from all different substances on Earth; they are also equal to the corpuscles emitted sometimes in large swarms by the Sun, which colliding with the Earth's atmosphere give rise to the phenomenon of the Northern Lights.

The electron has a negative electric charge and matter, being electrically neutral, must also contain a positive component. The investigation of the nature of the positive component gave rise to many discussions. At the beginning of this century, Thompson's theory was in vogue; this theory assumed the atom to be a sphere of positive electricity containing negative electrons, which when performing complicated movements would give rise to light phenomena. This representation was soon replaced by another, which is in some sense the antithesis of the previous one, due to Rutherford, who assumed the positive electricity of an atom to be concentrated in a corpuscle of dimensions much smaller than those of the atom, and called this corpuscle the {nucleus.} Rutherford came to this conclusion thanks to famous experiments based on the following principles. Radioactive bodies spontaneously emit alpha particles among their radiations, which are electrically positive corpuscles of such a high velocity that they can cross thin layers of matter, before being stopped by collisions against atoms. The alpha particle being charged with positive electricity is repelled when approaching the positive charge of an atom, thus deviating its trajectory; a strong deviation occurs only when the particle passes very close to the charge itself. The frequency with which deviations in the trajectories of the alpha particles are observed, naturally depends on the distribution of the positive charges: the higher the concentration of positive charge, the higher the probability of deviations at large angles. Comparing the results of the observations on the frequency of the deviations of the particles with theoretical predictions corresponding to several hypotheses on the distribution of positive electricity, Rutherford concluded that the best agreement between theory and experiment was obtained by assuming that the positive charge of the atom was all concentrated in its center. 

The atom is therefore constituted by a positive nucleus located in the center, around which electrons move. These electrons exist in sufficient number to neutralize the positive charge of the nucleus with their negative charge. This model of the atom is analogous to a tiny planetary system, in which the nucleus corresponds to the Sun and the electrons to the planets. If the nucleus has, for example, charge +7, seven electrons, each one of charge -1, are needed to neutralize it. This results in the enormous importance of the charge of the nucleus in fixing the properties of the atom, since the number of electrons that form it depends on this charge. If the nucleus has a charge +1, the neutral atom contains only one electron, and this is the hydrogen atom; if it has a charge +7 it is neutralized by 7 electrons, constituting the nitrogen atom; a nucleus of charge 92 with 92 electrons forms the uranium atom.

The model of the atom as a planetary system that we have just described is the one that is still supposed to be correct today, and has suggested for some time that the treatment of atomic physics problems could be made with the same methods of celestial mechanics following the success they have had in astronomy to explain the motion  of  planets. Experience, on the other hand, has shown that such a simple solution is not possible. The laws of classical mechanics have been obtained from experiments carried out partly on bodies of ordinary dimensions and partly directly on objects of very large dimensions, such as planets and stars. Pretending to apply without modification these same laws to a corpuscle of the dimensions of the atom undoubtedly constitutes a rather risky extrapolation and one should not be surprised if this leads to failure.

In the next lecture I will try to give an idea of the changes that have been introduced in the classical laws of mechanics in order to build a new mechanical theory suitable for dealing with atoms.


\vskip 1cm

\section*{Buenos Aires: Lecture II\\The concept of measurement and its criticism (August 6, 1934)}

\vskip 0.3 cm

\addcontentsline{toc}{subsection}{II: The concept of measurement and its criticism}

In my previous lecture I tried to give an idea of the anatomy of the atom, i.e., of the structural elements that constitute it. In this second general lecture, I will try to give an idea of the laws that govern the motion of the constituent elements of the atom and thus determine the physical and chemical properties of matter.

Summarizing what I said   last time, let me remind that we had come to the description of Rutherford model, according to which the atom is constituted by a positively charged central nucleus, in which almost all the mass is concentrated, and a number of electrons that move around it with a complicated motion. The number of electrons is determined by the magnitude of the charge of the nucleus: being the atom neutral, it must have as many electrons as needed to neutralize the positive charge of the nucleus. Then I introduced the problem of the investigation of the laws according to which the motion of the electrons  determining the properties of the atom occurs.

From the first attempts to quantitatively study the behavior of the atom, it was recognized that the ordinary laws of mechanics and electromagnetism could not be applied to it. The reason is extremely simple: it results from the general principles of electromagnetism that, when an electric charge describes a motion that is not rectilinear and uniform, it radiates electromagnetic energy; and the electrons in an atom describe precisely movements with acceleration. The phenomenon is analogous to the irradiation of radiotelegraphic waves, produced by an alternating current that flows through the antenna. 
So, oscillating electrons emit energy and the amplitude of their vibrations is rapidly extinguished because energy is lost, radiated in the form of electromagnetic waves.

According to classical electromagnetism, while traveling in an orbit an electron in an atom radiates energy and therefore its own energy decreases; this fact would translate into a continuous decrease of the radius of the orbit, after some time the electron would end up falling into the nucleus. It follows that if the same laws of mechanics and electrodynamics of macroscopic bodies were valid for the atom, we would not be able to understand how the atom could maintain its existence and how the electrons could continue to rotate around the nucleus without falling into it. We  then conclude  that the classical laws are not valid (as we had suspected), and therefore they should be modified. Here comes the problem of searching for the laws that must replace the old invalid ones.

The beginning of a solution to this problem, although provisional, appeared in 1913 and was due to the Danish physicist Niels Bohr, who based his research on the concept of the ``quanta'' of energy, introduced in physics in the study of thermal radiation.

By studying this problem\footnote{Fermi is speaking here of the problem of the so-called  ultraviolet catastrophe; it is not clear if there are omissions in the transcription.  Classical physics predicts that an ideal body at thermal equilibrium  emits radiation in all frequency ranges,  releasing an arbitrarily high amount of energy. This would cause all matter to  radiate
almost instantaneously all of its energy, which is clearly a paradox (note of the editors).} with the instruments of classical mechanics and physics, a disagreement was reached between theory and experiment, analogous to that found with classical mechanics applied to the atom. Numerous attempts were made to tweak the laws and try to overcome that inconvenience. Planck proposed a successful solution based on a hypothesis that must have seemed very strange and daring to the physicists of his time. The principle is as follows. In a mechanical oscillator, the energy of motion according to classical mechanics can vary continuously from zero to infinity; Planck, on the contrary, assumes that physical reality is different, and that the energy of an oscillator can only take integer, discontinuous,  values multiple of a certain energy, called the ``quantum'' of energy. He also conjectures that between the energy $\epsilon$ and the frequency $\nu$   a proportionality relation exists: $\epsilon = h \nu$, where $h$ is a universal constant independent of the  properties of the various bodies. This constant is called the Planck constant.

I believe that Planck himself, in the early days, did not believe to the letter in this hypothesis, but rather considered it as having a heuristic value, probably without a correspondence to physical reality. However, Planck was able to demonstrate that by assuming this granular structure of energy, the difficulty of interpreting the spectrum of incandescent bodies was resolved.

The second application of quantitative concepts came a few years later, that is, in 1905, and was due to Einstein, who used it to interpret the photoelectric effect. He argued according to Planck's ideas that the energy of light is not distributed throughout the light wave, but concentrated in energy granules, or in packets having each an energy content $\epsilon = h \nu$; these granules were called ``quanta of light.'' Such a conception, which represents in a certain sense a return to the corpuscular theory of light, allows us to understand how a very weak illumination can determine the expulsion of electrons from different bodies, provided that the frequency of light is large enough to deliver sufficient energy to an electron to overcome the attractive forces that bind it to matter, when the atoms constituting this matter absorb a ``quantum'' of energy.

The introduction of the notion of the quantum of light, which, as we have said, represents a return to corpuscular theory, seemed for a long time to be in direct contrast to the whole complex of phenomena that has led to  the wave theory. We will see at once how the contrast is only apparent and how modern theories allow us to give a unitary interpretation of the diffraction phenomena and of those in which energy changes occur by ``quanta''.

Let us now return to Bohr's theory on the structure of the atom and, for the sake of simplicity, let us refer to the easiest case of the hydrogen atom, consisting of a nucleus and of a single electron. The force that attracts the electron to the nucleus due to Coulomb's attraction is inversely proportional to the square of the distance and therefore obeys the same law of the gravitational attraction exerted between the planets and the Sun. If the same laws of classical mechanics regulating the motion of the planets were valid in the atom, the motion of an electron should be equal to that of a planet except for the difference in dimensions. The electron's orbit would therefore be an ellipse with the nucleus in one of its foci. If for simplicity we limit ourselves to circular orbits, we find that because of the different radii  they can have, the energy of the motion  could take continuous values, and the radii of the possible orbits could vary continuously from zero to infinity.

The fundamental hypothesis of Bohr's theory is to assume, contrary to this result of classical mechanics, that not all mechanically possible states are actually physically achievable, and that the only physically achievable states, called ``quantum states'', are a discrete succession of states selected with certain particular rules, indicated by Bohr, and which do not need to be specified here. Bohr also conjectures that when the atom radiates energy, it abruptly passes from one quantum state to a different one, emitting the difference in energy of the two states in the form of radiation of one single quantum of energy.

These ideas by Bohr allowed to give a quite satisfactory qualitative interpretation of a whole complex of properties, especially spectroscopic, of the various atoms. In no case, however, were these considered as a satisfactory solution to the problem of the atom, either because they were logically incoherent, or because they proved to be insufficient in many cases, for an exact quantitative description of the phenomena.

As a consequence, there was a series of attempts to determine on a more solid basis the physics and the mechanics of the atom. After many failures, an atomic mechanics has been built based on a total revision of the same fundamental kinematic concepts and which can now be considered as a satisfactory solution to almost all of the atomic problems, except those related to the structural properties of the electron or the nucleus.

Without going into the mathematical details of this new atomic mechanics, I will try to explain which concepts have served as the basis for its construction.

With this aim, as in the theory of relativity, the starting point was the criticism of intuitive concepts that, under a rigorous examination of the concrete possibility of controlling them, are insufficiently defined. In the case of the theory of relativity, the starting point was the critique of the concept of simultaneity of two events, which led Einstein to recognize that simultaneity has no absolute value, but is relative to the system in which observations are made, thus implying that two events that are simultaneous for one observer are not simultaneous for a different one.

In the construction of atomic mechanics, a criticism of the kinematic concepts of position and velocity of a material point was necessary. This criticism was developed by discussing the concrete possibility of measuring  these two properties together. It is necessary to understand that this criticism is not referred to the technical difficulties of the measurements. The main question here is the analysis of the limitations in the precision of the measures, 
which are inherent not to imperfections of the instruments or the methods, but to the very nature of the phenomenon under examination. From this analysis it turned out that in the problems of atomic physics, it is not possible to avoid the disturbances that the   measurement action itself exerts on the phenomenon under examination. The criticism culminated in the so-called Heisenberg principle of indetermination, according to which there is, at least in principle, no limitation to the exact measurement of the position or speed separately, while a limitation exists for their simultaneous determination. In fact, a measure of the exact position necessarily produces a very large and uncontrollable alteration of the velocity, and a measure of the velocity produces an uncontrollable alteration of the position.

The new atomic mechanics is often called wave mechanics, because one of its most significant mathematical forms consists in associating the movement of a material point with a wave system   analogous to the wave system associated with the motion of a quantum of light. Thus, from a historical point of view, wave mechanics came through de Broglie and Schr\"odinger starting directly from this analogy and only later it could be shown that the wave scheme was mathematically equivalent to that proposed by Heisenberg, based on a criticism of the principle of measure.

In the new mechanics the ``wave-corpuscle'' dualism has found a harmonic interpretation since the two conceptions, both related to the case of light radiation and to the dynamics of a point, can be presented as two different aspects of the same phenomenon.

In the new mechanics, finally, there is the possibility of a quantitatively exact and logically unitary description of the entire complex of spectroscopic and chemical phenomenology, of which the early Bohr concepts allowed only qualitative interpretations.

Of course, the validity of this new theory has also its limitations. Such limits are found when phenomena are considered in which the structure of the electron or that of the nucleus intervenes, that is to say when these bodies can not be considered as point-like, and therefore their properties related to extension or internal complexity must be taken into account.

This opens a new series of problems whose main objective is to investigate the structure of the atomic nucleus.

In the next lecture, we will review some of the main results obtained to date in this new field of physics.


\vskip 1 cm

\section*{Buenos Aires: Lecture III\\The fundamental elements of nuclear structure, Part 1}

\vskip 0.3 cm

\addcontentsline{toc}{subsection}{III: The fundamental elements of nuclear structure, Part 1}

In the physics of the atom, the structure of the nucleus is usually ignored, since this is considered as a material point, given its extreme smallness. This description, that is sufficient  for most atomic problems, is obviously incomplete. The most convincing demonstration of this fact comes from radioactive phenomena, 
which have been known for almost fifty years, and consist, we know, in a spontaneous disintegration of the nuclei of the heaviest elements.

The data available for the various species of nuclei are, first of all, their electric charge, the value of which coincides with the atomic number when the charge of the electron is taken as unity, and their mass number. The latter always has values  close to integers, as measured with the ordinary unit with which the atomic weights are measured, taking into account of course the existence of the various isotopes.

These facts have allowed us to suppose for a long time that the various nuclei are made up of different aggregates of the same fundamental elements. Until a few years ago, these fundamental elements of the nuclear structure were thought to be the two simplest particles hitherto known: the electron (with electric charge -1 and negligibly small atomic weight) and the nucleus of hydrogen, or proton (with electric charge +1, and atomic weight close to 1). By adding a convenient number of electrons and protons, it was always possible to obtain the necessary values for the electric charge and for the atomic weight of the nucleus. Thus, for example, the nitrogen nucleus, that has an electric charge +7 and an atomic weight 14, can be thought as made up of an aggregate of 14 protons and 7 electrons.

The fundamental elements that we can use to build a general model of the nuclear structure have increased today thanks to the discovery of two new corpuscles:\footnote{Throughout his lectures and especially in this lecture, Fermi uses frequently the term {\em ``corp\'usculo''} (corpuscle) and only rarely the term {\em ``particula''}  (particle), however with the same meaning (note of the editors).} the neutron that has zero electric charge and atomic weight close to 1, and the positive electron or ``positron'' that differs from the common electron only in the sign of its electric charge. Thanks to these discoveries, the ideas about the structure of the nucleus have currently undergone a considerable revision so that a new scheme, according to which the fundamental constituents of all nuclei are protons and neutrons, is currently preferred to the previous one (a construction based only on protons and electrons). The reasons for preferring this new scheme to the original one derives mainly from theoretical difficulties that make it hard to understand how light corpuscles can be maintained in a region of extremely small dimensions such as the nucleus. It results from quantum mechanics that, while there are no difficulties in understanding that relatively heavy corpuscles such as the proton and the neutron stay in the nucleus, supposing  that light corpuscles such as the electron or the positron can stay as well in the nucleus would require deep modifications of the fundamental laws.

Another type of difficulty is linked to  the fact that, if certain fundamental concepts of wave mechanics are also applicable in the nuclear field, the analysis of the nuclear spectra, for example, that of nitrogen, establishes that the number of elementary corpuscles contained in the nucleus of this element must be even. We have just said that if nuclei were formed by protons and electrons, the nitrogen nucleus should have 14 protons and 7 electrons, that is to say, in total, an odd number of corpuscles. On the other hand, if we assume that nuclei are formed by protons and neutrons, a nucleus of charge +7 and atomic weight 14 is made by an aggregate of 7 protons and 7 neutrons, and therefore the number of constituent corpuscles is even. This evidence that we have illustrated with the example of nitrogen is present in other cases in which it is found that assuming a nuclear structure based on neutrons and protons solves the difficulties that would result from assuming nuclei made up of protons and electrons.

The discovery of the neutron was made in 1932, and is due to a series of works by the German Bothe, the French Curie and Joliot and the English Chadwick.\footnote{Chadwick was awarded the Nobel prize in physics in 1935 ``for the discovery of the neutron'' (note of the editors).} Neutrons are released spontaneously and are emitted with a considerable energy by the nuclei of the light elements and particularly by beryllium, under the action of bombardment with $\alpha$ particles. Their characteristic properties and in particular their high penetrating power depend essentially on the absence of electric charge. Indeed the deceleration that a charged particle undergoes, for example, as it passes through matter, depends on the fact that as the particle approaches the electrons of the atoms, it exerts on them electrical forces that set them in motion, thus determining the transfer to them of a part of its energy, in such a way that the particle rapidly loses speed until it stops. On the other hand, the neutron,  having zero electric charge, only interacts with other corpuscles only when it approaches them at very small distances, and therefore the deceleration it undergoes when passing through matter is quite small, so that its penetrating power turns out to be a few thousand times higher than for the $\alpha$ particle.

What is  observed in practice are collisions between neutrons and the nuclei of the matter  they go through. On the contrary, collisions between a neutron and an electron are extremely rare, a fact that may seem strange given that electrons actually exist in considerably greater numbers than nuclei. The reasons for this, which would be difficult to understand with the concepts of classical mechanics, are found instead by studying the collision phenomenon based on wave mechanics. This leads to establish that the probability of collision between a neutron and a corpuscle is approximately proportional, neglecting other differences, to the square of the reduced mass of the system of the neutron and the colliding corpuscle. This reduced mass in the collision between a neutron and a heavy nucleus is close to the neutron mass, which is in turn similar to the proton mass and we define as 1, while in the collision against an electron it is roughly equal to the mass of the electron itself, that is, about 1/1800 of the mass of the proton or of the neutron. It follows, therefore, that the probability ratio for collision against a nucleus or against an electron is of the order of 1000~000, which certainly justifies the rarity of collisions with electrons.

In the same year 1932 in which the neutron was discovered, the existence of the new elementary particle that we have already mentioned, the positron, was also established.

The existence of positrons was recognized for the first time in the secondary phenomena that accompany cosmic radiation. Although the nature of this cosmic radiation is still being discussed today, it has been experimentally proven that it contains corpuscles of energies up to a thousand times higher than those obtainable in the laboratory. The analysis of the sign of the electric charge of these particles can be carried out by examining the direction in which they deviate when they cross a magnetic field. It was precisely by photographing the trajectory of corpuscles produced by penetrating radiation in a Wilson chamber\footnote{The Wilson chamber \cite{intropart}, or cloud chamber,   invented by C.T.R. Wilson at the beginning of the
twentieth century, was an instrument for reconstructing the trajectories of charged particles. The
instrument is a container with a glass window, filled with air and saturated water vapor; the volume
could be suddenly expanded, bringing the vapor to a supersaturated (metastable) state. A charged particle crossing the chamber produces ions, which act as seeds for the generation of droplets
along the trajectory. One can record the trajectory by taking a photographic picture. If the chamber
is immersed in a magnetic field, momentum and charge can be measured by the curvature (note of the editors).} that Anderson in California found for the first time indications of the existence of corpuscles whose tracks, in all respects appearing as electron tracks, were deflected by a magnetic field in the opposite direction to that of the electrons, thus indicating that the charge of those corpuscles was positive and not negative.

The most beautiful experiments in this field are due to the English Blackett and the Italian Occhialini. Using an ingenious device to be able to photograph with a Wilson chamber a large number of corpuscles produced by penetrating radiation, these physicists could photograph extremely complicated interaction and disintegration   phenomena, probably explicable by the great energy of cosmic rays, in which from the same  center it is possible to observe the radiation of a few tens of high-speed corpuscles. If the phenomenon is observed by applying a magnetic field to the Wilson chamber, it can be recognized that some of these tracks belong to negative electrons, while others deviate in the opposite direction and must be attributed to positive electrons.

After the discovery of the positron, the presence of this particle has also been observed in some other circumstances reproducible in the laboratory. In particular, it was discovered that, under the action of very hard $\gamma$ rays in high atomic weight materials, positive and a negative electron pairs, that are emitted from the same point, are produced.

The discovery of the positive electron can also be considered interesting because, prior to experiment, theory had already made it possible to conjecture its existence.

We will see in the next lecture what is the theoretical importance of the discovery of the positive electron and what are the properties of this new particle.

\vskip 1 cm

\section*{Buenos Aires:  Lecture IV\\The fundamental elements of nuclear structure, Part 2}

\vskip 0.3 cm

\addcontentsline{toc}{subsection}{IV: The fundamental elements of nuclear structure, Part 2}

In the previous lecture we have examined the  discovery of the positive electron. This experimental result has also been a great success for  theory, since some years before its discovery, the possible existence of positive electrons had been predicted based on theoretical considerations.

Such a prediction was due to the English physicist Dirac; he reached it trying to build up a wave mechanical description of the electron consistent with the theory of relativity. The theory that he thus constructed seemed immediately very interesting, especially because without the need for ad hoc hypotheses, it
accounted for the existence of a magnetic moment of the electron, which is necessary for the interpretation of the energy spectrum of photons emitted in atomic transitions.

Despite this success, Dirac's theory had a serious drawback, since, in addition to predicting
the energy levels of the physically possible electron states, it allowed other states whose kinetic energy turned out to have a negative value instead of a positive one and whose interpretation seemed mysterious at the beginning. For example, for an electron not subjected to forces and with a momentum $p,$ the kinetic energy in Dirac's theory has the following expression:
\[ L = \pm \sqrt{m^2c^4 + c^2 p^2} \, . \]
The states corresponding to the positive sign coincide with the ordinary expression of the kinetic energy of the theory of relativity, as it also includes the term $mc^2,$ representing the energy equivalent to the  electron mass. However, no possible interpretation was found for the negative sign of the radical, since it  can not describe the behavior of a material point.

To solve this problem, Dirac made a hypothesis that seemed at first quite strange: he actually supposed  that all the states corresponding to a negative kinetic energy should be considered occupied in the sense of the Pauli exclusion principle, which gave a direct explanation of why it is impossible, at least under normal conditions, that an electron goes from a state of positive energy to one of these abnormal states. According to this hypothesis, the physically empty space should be considered as corresponding to the case in which all infinite states with negative energy are occupied by one electron each, while those with positive energy are all free.

\begin{figure}[h]
\begin{center}
\includegraphics[width=.7\textwidth]{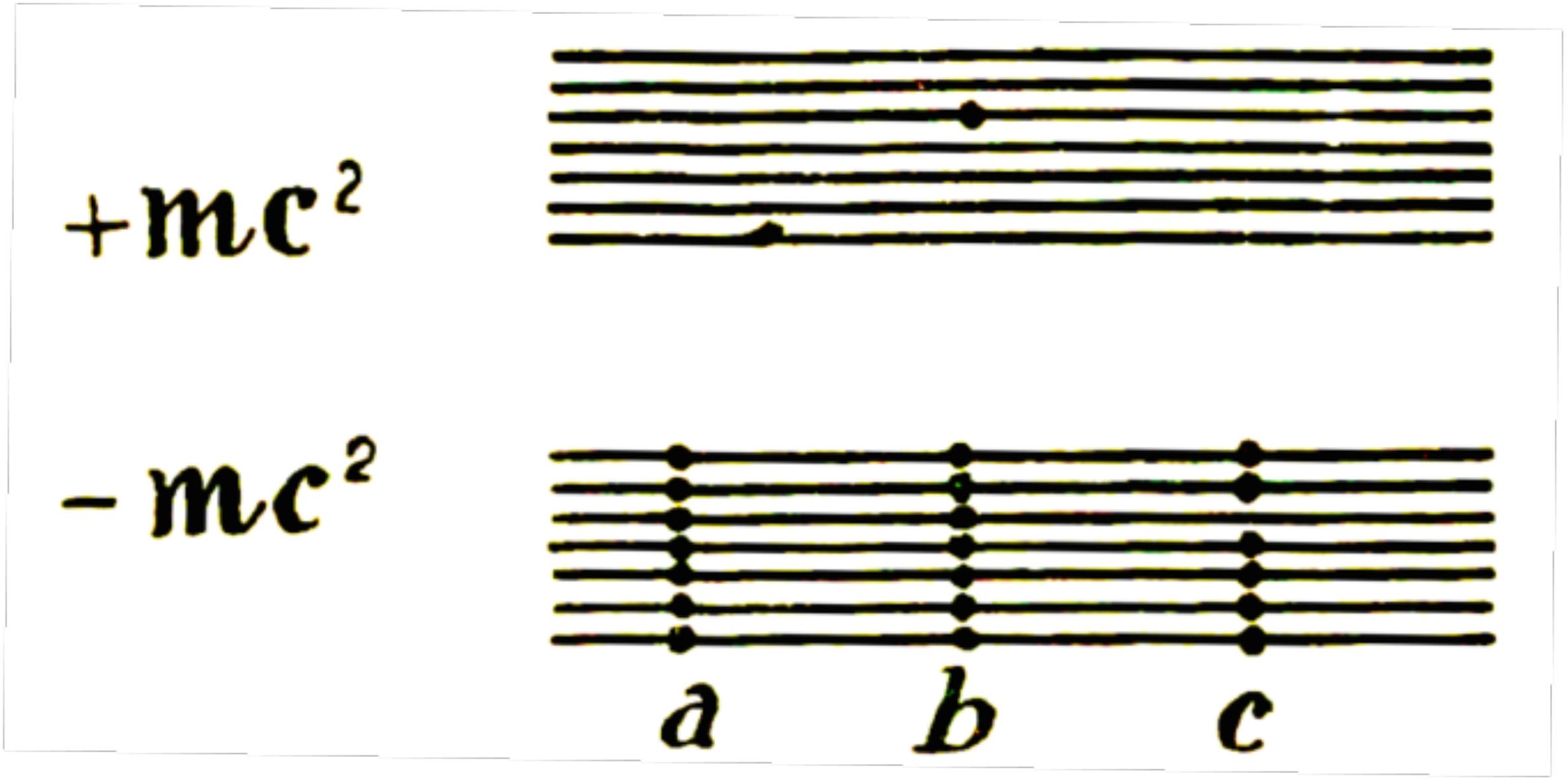} Fig. 1
\end{center}
\end{figure}

In Fig. 1, the states are schematically represented by horizontal lines; those with positive energy extend above $+mc^2$ and the negative ones below $- mc^2.$ Indicating with a dot that the state is occupied by an electron, the empty space is schematized by the situation represented by $a$. On the other hand, in case there is a single electron, it is represented by $b:$ all negative states are occupied as also a positive one. Conversely, case $c$ represents the situation in which all positive states are free and negative states are all occupied except for one, or in other words, there is a hole in the negative distribution. It can be realized that the physical properties corresponding to this hole are identical to those of an electrically charged corpuscle of opposite charge to that of the electron, that is, that the hole behaves like an electrically positive corpuscle because it corresponds to the absence of a negative electron in the place of the ``hole''. At first Dirac thought that these holes behaving like electrically positive corpuscles should be identified with protons; however this idea had to be abandoned later, because it was shown that even taking into account the electrical interaction between all the electrons, it always turned out that a normal electron and a hole necessarily have the same mass, whereas the electron and the proton have very different mass.

The experimental discovery of the positive electron has simply allowed us to interpret these new corpuscles as the holes in Dirac's theory.

Other important properties of the positive electron or ``positron'' find an easy interpretation if the theoretical scheme already exposed is accepted. When an electron and a positron\footnote{Here the Spanish transcription says, wrongly, proton instead of positron (note of the editors).} get close, it can happen that they annihilate each other, neutralizing their opposite electrical charges and transforming their mass into radiant energy. In Dirac's theory this phenomenon is interpreted as a transition in which the electron passes from the positive state in which it is initially found to the hole-state corresponding to the positron. In this way, the destruction of both corpuscles is evidently produced. To this process, that we can call dematerialization process, a materialization process is opposed in which an electron passes from an occupied state of negative energy to one of the higher states, thus forming an ordinary electron and a positron that corresponds to the remaining hole in the negative state where the electron comes from. This materialization process with the creation of an electron-positron pair has been observed by irradiating  substances of high atomic weight  with very hard $\gamma$ rays. In fact, the absorption of a $\gamma$ quantum and the creation of an electron-positron pair with the described mechanism sometimes happen in the neighborhood of the nucleus.

Turning now to the properties of the nucleus, we will briefly expose some theoretical views on radioactive phenomena. These phenomena consist, as is well known, in the spontaneous disintegration of some nuclei, particularly heavy elements, which allow the emission of $\alpha$ or $\beta$ particles   with high speed, with the subsequent transmutation into new nuclei. The emission of these particles is often accompanied by electromagnetic irradiation, i.e., $\gamma$ rays. As an $\alpha$ particle is   a nucleus of $^4$He (it has a charge +2), its emission determines a decrease of two units of the electrical charge of the nucleus that expels it; the residual element therefore will be displaced by two positions  to the left of the original element in the periodic table of Mendeleev. On the contrary, the emission of a $\beta$  particle  (that is an electron and has electric charge -1) results in the displacement of a position forward, since it increases the nuclear charge by one unit.

Applying the principles of wave mechanics to nuclear problems it has been quite easy to construct a satisfactory theory for the emission of $\alpha$  particles. In this theory, which is due to the Russian physicist Gamow, it is admitted that when an $\alpha$  particle is outside the nucleus, it is rejected by it in accordance with Coulomb's law. On the other hand, it is supposed that the attractive forces needed to justify the fact that the particle can be kept inside the nucleus for a long time act inside the nucleus. In other words, it is assumed that the force acting on the $\alpha$ particle depends on its distance $r$   to the center of the nucleus according to a law of the type represented graphically in Fig. 2, in which $\rho$ represents the radius of the nucleus. The potential energy has a value W$_0$ at a distance from the nucleus approximately equal to $\rho$. According to classical mechanics, an $\alpha$ particle initially located inside the nucleus, with an energy positive but smaller than W$_0$, could never leave it, not possessing enough energy to cross the barrier of high potential energy that surrounds the nucleus. According to wave mechanics it is found, on the contrary, that this impossibility of escaping  is not absolute and that instead there is a finite probability that the particle crosses the potential barrier leaving the nucleus. The probability of this process becomes smaller the higher and wider the barrier to cross.

\begin{figure}[h]
\begin{center}
\includegraphics[width=.65\textwidth]{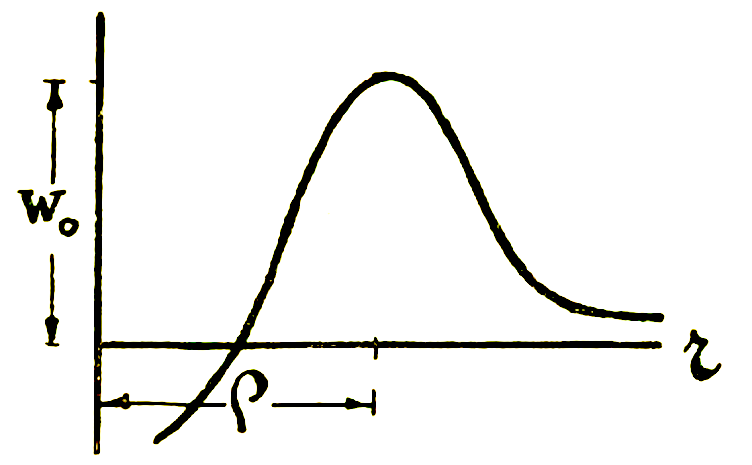} Fig. 2
\end{center}
\end{figure}

The possibility of this passage is closely linked to the undulatory nature of particles, postulated by wave mechanics and according to which even in the regions where with classical mechanics the kinetic energy would be negative and hence the particle   could not penetrate, the tails of the wave representing the particle can enter, positioning the particle itself beyond the region forbidden according to ordinary mechanics.\footnote{This is today known as ``tunnel effect'' (note of the editors).}

Based on these concepts, Gamow was able to provide a satisfactory interpretation of most of the phenomena related to the emission of $\alpha$ particles and in particular of the relationship that exists between the 
mean lifetime and the energy with which the particles are emitted. His theory also gives perfect account of the enormous variations in the magnitude of the lifetime that, for various radioactive substances, ranges from very small fractions of a second to many millions of years.

While the theory of the emission of $\alpha$  particles can currently be considered  quite satisfactory, the same can not be said for the theory of the emission of $\beta$ particles, and this is so for two reasons. First: we have seen the difficulties in admitting the existence of electrons inside the nucleus, so that it seems difficult, at first glance,  to understand  how they can get out of it. {{Second: while $\alpha$ particles, at least in typical cases, all leave the nucleus with the same energy, the $\beta$ particles leave the nucleus with a continuous distribution of energy;}} in other words, it is found that in the $\beta$ transitions the starting nucleus and the final product have both well-defined energies, while the emitted $\beta$  particle can have different energies, which seems to be in conflict with the principle of energy conservation. And indeed, very authoritative physicists and among them Bohr (first of all) have expressed the opinion that this may be a case in which the principle of conservation of energy is not fulfilled. Of course energy balance could be restored due to the existence of a new form of energy that escapes the current observation methods. This should possess a very high penetrating power since it does not seem to be detected by the devices that are ordinarily used. A hypothesis of this kind has been supported by Pauli, who has thought about the possibility of a new elementary corpuscle of mass comparable to that of the electron and electrically neutral: this hypothetical corpuscle has been given the name of neutrino. The difficulty of conserving energy in the emission of $\beta$ rays would be eliminated, assuming  in a $\beta$ process 
the simultaneous emission of an electron and a neutrino  between which the energy  released in the process could be distributed in various ways, thus granting the possibility that electrons leave the nucleus  with different energies. On the other hand the neutrino, given its electrical neutrality and its tiny mass, could practically escape any current observation possibility.\footnote{Fermi's theory of $\beta$ decay is currently accepted today. The neutron $n$ transforms into a proton $p$ through the process $n \rightarrow p e^- \bar{\nu}_e$, the  negative electron $e^-$ being the $\beta$ ray, and  $\bar{\nu}_e$ the (anti)neutrino (note of the editors).}

A couple of words should be added on  the other difficulty we had mentioned. Electrons leaving the nucleus would be interpreted in the most obvious way, assuming that they were initially contained in it, but since we have seen that  this assumption creates serious difficulties, another hypothesis can also be put forward, namely that the electron and possibly also the neutrino are  the result of the $\beta$ transformation. This would consist
in a process in which a neutron transforms into a proton, thus obtaining an increase of one unit in the nuclear charge, while the formation of an electron occurs simultaneously in such a way that the conservation of the electrical charge is restored. Assuming the creation of another additional particle is not an entirely new hypothesis, since there is already an example in radiation theory: when an excited atom makes a transition into a lower energy state emitting radiation, a quantum of light, which can be considered as being created upon emission and not as pre-existing inside the excited atom, leaves it.

One can thus try to construct a theory of $\beta$  rays analogous to the theory of irradiation of light quanta. Although   such a theory is qualitatively possible according to experimental facts, it is clear that it has to be very cautiously considered  at the moment, since it implies the existence of the neutrino, that is to say, of a particle whose existence is currently lacking any direct evidence.

\vskip 1 cm

\section*{Buenos Aires:  Lecture V\\The artificial disintegration of the nucleus}

\vskip 0.3 cm

\addcontentsline{toc}{subsection}{V: The artificial disintegration of the nucleus}

The topic that I want to address in this last lecture concerns what can be done artificially with the atomic nucleus. While it is relatively easy to produce modifications in the external electronic structure of the atom, altering the organization of the nucleus is a much more complicated problem for two reasons: first, because the nucleus is very small and therefore more difficult to struck by  bombardment, and second because the corpuscles of the nucleus are bound with much greater energies and therefore the projectiles must have a very high energy to be able to produce a break. The first projectiles that have demonstrated to be able to break the nucleus were the $\alpha$ particles that are spontaneously produced by radioactive atoms. In this field, the first successful experiment   was carried out by Lord Rutherford in 1919; one of the best-known cases of this process was the disintegration of nitrogen. The nitrogen nucleus has charge 7 and atomic weight 14, which is indicated with $^{14}_7$N. When the nucleus is hit by an $\alpha$ particle, it absorbs it and immediately after emits a proton. The $\alpha$ particle is a nucleus of helium of charge 2 and atomic weight 4 ($^{4}_2$He); therefore, when it collides with a nucleus of nitrogen, a nucleus of charge 9 and of atomic weight 18 is formed by fusion. Then, this nucleus immediately emits a proton, i.e. a hydrogen nucleus ($^{1}_1$H). The resulting nucleus then has charge 8 (it is therefore oxygen, O) and atomic weight 17, that is, $^{17}_8$O. The common O has atomic weight 16, but there is a rare isotope that has weight 17; the process is obviously a true transformation of elements in which N is transformed into O. This nuclear reaction has been particularly well studied, since one can observe  in a Wilson chamber the path of the $\alpha$ particle as well as that of the proton and of the nucleus of O, the latter acquiring a certain speed due to the shock it receives.

In the first disintegration experiments, the bombardment was obtained by means of $\alpha$ particles, that is, by means of a  projectile spontaneously emitted by radioactive substances. In recent years, speeds comparable to those of $\alpha$ particles have been obtained for nuclei, and therefore it has also been possible to use other projectiles, protons in particular. The first experiments of this type were due to Cockcroft and Walton,  who built a discharge tube in which hydrogen rays, which are protons, are sent through potential differences that reach almost a million volt. By shooting these protons against the nucleus of some substances, particularly lithium (Li), it has been possible to observe the disintegration of this element. Lithium has two isotopes, one of atomic weight 6 and the other weighting 7; the one that is affected by the interaction is the one of weight 7. When a proton strikes one such nucleus, it is incorporated by lithium, and an aggregate is formed with mass 8 and charge 4, that is, twice as much as the mass and charge of  the $\alpha$ particle. This aggregate is unstable and immediately breaks down into two $\alpha$ particles according to the following nuclear reaction:
\[ {\rm{^{7}_3{\rm Li} + ^{1}_1{H} \rightarrow 2 ^{4}_2{\rm He}}}  \, \]

Also in this case the phenomenon can be examined in a Wilson chamber, where the track of the proton acting as a projectile and the two tracks of the $\alpha$ particles, approximately in opposite directions to each other, are observed.

Protons are in a sense more effective in producing nuclear decays than $\alpha$ particles, at least with the same energy. Indeed, the proton has charge 1, $\alpha$ particles have charge 2, and therefore when these two projectiles approach a nucleus, the electrostatic repulsion for  an $\alpha$ particle is double  than for a proton. This reason, and another one which can be deduced from wave mechanics and depends on the fact that the proton has a smaller mass than the $\alpha$ particle, explain that, at least in the case of Li, which is one of the lightest elements, disintegration occurs even with relatively very small energies of the bombarding particle. Appreciable effects have been obtained even with protons artificially accelerated through potentials smaller than 20\,000 volt. Neither with $\alpha$  particles nor with protons it has been possible to obtain effects larger than 
those with light elements: the electrostatic repulsion is too strong and the particles fail to reach the nucleus. This difficulty does not show up when the bombardment is carried out with neutrons, which do not suffer electrostatic repulsion and can therefore reach the nuclei of the heaviest elements.

Until this year it was believed that the atom formed in these artificial decays was always necessarily a stable atom; in January, however,   the French physicists Joliot and his wife Ir\`ene Curie, daughter of the discoverer of radium, 
announced the discovery of  three cases in which the products obtained by bombardment with $\alpha$ particles are not stable, but subsequently disintegrate behaving like naturally radioactive bodies. In their experiments, these physicists used  $\alpha$ particles, {{which a priori restricts the possibility of producing an effect only to very light elements.}}

In the experiments we have carried out in Rome, we planned to use neutrons to make the bombardment. Neutrons, being as  said electrically neutral, can collide even with very heavy nuclei since they do not undergo electrical repulsion. I will briefly describe how the experiment is performed. For this experiment we need a neutron source, the substance to be bombarded, and an apparatus to reveal the possible artificial radioactivity produced by the substance. 

The source is extremely simple. It is based on the fact that beryllium, under the action of the bombardment by $\alpha$ particles, spontaneously emits neutrons. Beryllium powder and a radium emitter are introduced into a glass tube that is  then sealed. The particles emitted by the radium emanation and by its decay products collide with the beryllium grains and activate the emission of neutrons. With the source that we were able to use, the number of neutrons fluctuated around one million per second; this is not a large number when compared to the number of particles produced by radium, in which a gram emits a few tens of millions of particles per second. This lower intensity is compensated by the fact that neutrons are not repelled; when a neutron is directed  towards a nucleus it collides with it, while under the same conditions the probability that an $\alpha$ particle produces an effect is very small, becoming practically negligible for heavy nuclei.

The apparatus for detecting a possible artificial radioactivity must be sensitive to the rays that emerge from the possibly formed radioactive substance.

The enormous sensitivity of the devices that can be built to reveal $\beta$ rays allows identifying the production of quantities of substances that would escape any other current method of observation. The apparatus for revealing the presence of electrons is a Geiger and M\"uller wire counter. It consists of a metallic tube at the ends of which there are two insulating caps connected by  a metallic wire, generally made of aluminium, that extends along the axis of the cylinder. Between the tube and the wire a potential difference is established, adjusted so that it is by little insufficient to produce the electric discharge. If an electron enters this space, it produces a small ionization that is sufficient to cause a very short-lived discharge, which is quickly extinguished. But during the time that this discharge lasts, it produces an alteration in the potential of the wire, which, suitably amplified, activates a counter that records the number of electrons that have entered the apparatus. Of course, in our case, the walls had to be thin enough for electrons to enter the chamber; they were made of aluminium 0.1 to 0.2 mm thick.

In order to examine a substance, Fe for example, a cylinder of suitable size is taken so that it can wrap up  the counter. First, the source is placed on the cylinder and left for a certain time during which iron remains exposed to the bombardment of the neutrons that come out of the source; some of the nuclei reached by the neutrons are modified and become radioactive. After the neutron source is removed, these modified, radioactive, nuclei, from time to time disintegrate, emitting an electron. If the counter is now inserted in the Fe cylinder, these electrons are registered by the counter and their number is recorded as a function of time. It is thus found that the frequency of disintegration decreases and reduces to one half of the initial one in a time characteristic of the bombarded element, which for example, in the case of iron, is equal to two and a half hours.

With this method we have tried to examine as many elements as possible, up to about sixty, and it has been found that a large number of them (about two thirds, that is to say about forty) produce radioactive effects. This large percentage of elements possibly subject to activation undoubtedly depends on the fact that neutrons are not repelled by the electric charge of the nucleus, and therefore can act even on heavy elements. As a confirmation of this, we have found that the percentage giving a positive effect is more or less the same for light as for heavy elements.

Let us discuss now the mechanism of the phenomenon and the experiments performed to analyze it. Above all, it must be verified that the phenomenon, like all nuclear phenomena, is independent of the type  of chemical bond. Thus, for example, Fe can be subjected to bombardment as an element or using its compounds, and the same phenomenon is always found with the same intensity (naturally taking into account the different Fe content) and with the same period;\footnote{Throughout these lectures, the half-life of the radioactive decay is reported by the transcribers as the ``evolution period'' or ``period'' (note of the editors).} the disintegration time is a characteristic constant of the bombarded element and it takes different values from element to element. The F activity is halved in 9 seconds, while for other elements the time is a few minutes, for others a few hours or a few days; the longest periods found so far correspond to Cl and S and are about 13 days.

The most probable scheme according to which decays occur is suggested by the already known cases of nuclear decay, where we know that  very often the bombarding particle is captured by the nucleus, which then emits a new particle. The three main schemes according to which the phenomenon occurs could therefore be the following:
\begin{enumerate}
\item Neutron absorption and emission of an $\alpha$ particle.
\item Neutron absorption only.
\item Other mechanisms (which can not be excluded). For example, there may be cases in which the neutron is not absorbed or where other particles than those mentioned are emitted. 
\end{enumerate}

If the nucleus absorbs a neutron and emits an $\alpha$ particle, its electric charge decreases by 2 units and therefore the radioactive nucleus formed is moved two places to the left with respect to the original element in the periodic table. If, on the contrary, a proton is emitted, a shift by only one position occurs, while in the third case the radioactive product formed turns out to be an isotope of the original element.

Even without making too specific hypotheses about the mechanism of the phenomenon, it can be reasonably assumed that the radioactive element has an atomic number close to that of the bombarded element. In the case of  Fe, for example,  we can limit the investigation to the elements close to it in the periodic system, that is, to Cr, Mn, Fe, Co and Ni. In order to decide among these possibilities, a chemical analysis of the formed element can not be carried out directly because its quantity is so small that it can not be evidenced with any of the ordinary methods of chemical analysis; it is actually necessary to add a radioactively inert isotope to the radioactive element possibly formed, {{which serves to extract it. A certain amount of metal is subjected to neutron bombardment and then dissolved in nitric acid; small amounts (a few milligrams) of the salts of all the  elements investigated (Cr, Mn, Co, Ni) are added to the nitrate solution;}} with the standard methods of chemical analysis these elements are separated again from each other and from Fe, and the five samples are successively placed next to the counter to know which of them is  radioactive. In this case, for example, it was found that the activity continues in  Mn, indicating that Fe has been transformed into a radioactive isotope of Mn, from which it would appear that in this case the radioactive product has an atomic number one unit lower than that of the original element and therefore a proton has been emitted. In other cases, it has been found that the atomic number decreases by two units, which would indicate the emission of an $\alpha$ particle. There have been cases, finally, in which the radioactive element is an isotope of the bombarded element, which could be interpreted  by assuming that the neutron has simply been absorbed without subsequent emission of charged particles.

The difficulty of performing a chemical analysis of the type described is in many cases due to the short time available to do it, which is sometimes 2 to 3 minutes. For this reason, chemical analyses of active products have been performed in about fifteen cases and it is sometimes found that the same radioactive element can be formed in different ways. For example, radioactive Mn that is formed by bombarding Fe, with a period of two and a half hours, is also obtained bombarding Co and Mn; analogously an isotope of V is formed, having a period of about 4 minutes, when bombarding V, Cr or Mn.

Finally,  let me come to  the activation of uranium (U), the heaviest of the known elements. The investigation is complicated in this case because U is spontaneously radioactive and its natural radioactivity would mask any further effects due to neutrons. To carry out the experiment, one can take advantage of the fact that  U only emits $\alpha$ particles, which are not registered by the counter because they have insufficient penetrating power to enter into it. But since U is always mixed with its decay products, some of which emit $\beta$ particles, chemical operations are necessary to eliminate them. After purification,  at least for some time, no spontaneous emission of $\beta$ particles occurs, and the activity is slowly restored as the separate decay products are formed again. After this preliminary treatment, the U subjected to neutron bombardment shows a relatively intense activity with a decay time very different from those corresponding to natural radioactive bodies. The phenomenon is rather complicated, because the decrease in the activity produced by the neutrons is not simply exponential, but is rather the superposition of several exponentials with different periods; one finds at least 4 components (10 seconds, 40 seconds, 13 minutes and 100 minutes), and possibly some longer. This indicates a greater complexity of the phenomenon, probably due to the fact that some successive transformations take place. Moreover, in the case of U it is interesting to chemically investigate the nature of the active elements formed. We have especially devoted ourselves to the 13-minute period element which is the most convenient for investigation and we have tried to establish whether this was an isotope of the U itself or of any of the lower atomic number elements ($_{91}$Pa, $_{90}$Th, $_{89}$Ac, $_{88}$Rd). Here, too, it is necessary to proceed in a slightly different way from that described for the common elements, because many of the elements with an atomic weight close to that of U are very rare and the quantities available are not large, but they can be revealed by radioactivity techniques. To investigate whether the element living 13 minutes is an isotope of one of these elements, the two materials are mixed and it is then searched for chemical reactions capable of separating them at least in part. The existence of one of such reactions should be considered as a proof that the two elements are not isotopes. The experiments carried out in this investigation seem to indicate that the element produced is not an isotope of U, nor of the elements that precede it within at least ten boxes of the periodic table; it has therefore been thought that it could have an atomic number greater than 92, in which case  an element that does not exist in nature would be  produced due to the effect of neutron bombardment, and  after an ephemeral life of few minutes it would again be destroyed. Clearly, before these results can be confirmed, it will be necessary to perform a series of control tests and to fully examine the set of  chemical properties of the new element.

Finally, let me add that, according to press news, it  seems that these days a Bohemian scientist\footnote{The alleged discovery of the element, now discredited, was made by the Czech scientist Odolen Koblic (note of the editors).} has found an element that he suspects to have atomic number 93 in the pitchblende from which the U is extracted. If this news were confirmed it would lead to a new possibility of obtaining element 93, perhaps in an isotopic form of that obtained with neutron bombardment.

\newpage

\let\oldthefootnote=\thefootnote

\newcommand{\astfootnote}[1]{
\setcounter{footnote}{0}
\renewcommand{\thefootnote}{\fnsymbol{footnote}}
\footnote{#1}
}

\section*{Experiments on artificial radioactivity\astfootnote{Spanish transcription by Dr.  A. Broglia and Eng. C. Vercellio.}\\(C\'ordoba, August 11, 1934)}


\setcounter{footnote}{10}

\vskip 0.3 cm

\addcontentsline{toc}{section}{Enrico Fermi's lecture in C\'ordoba (August 11)}

\addcontentsline{toc}{subsection}{Experiments on artificial radioactivity}


On August 11, 1934, at 18:00, in the Faculty of Exact, Physical and Natural Sciences of the University of C\'ordoba, the illustrious Professor Dr. Enrico Fermi, academician\footnote{Enrico Fermi was appointed in 1929 a member of the Reale Accademia d'Italia (note of the editors).}
 of Italy, has clearly and brilliantly presented a synthesis of the latest research carried out in the field of nuclear physics.

Professor Fermi has started his lecture reminding the phenomenon of the natural decay of radioactive elements that, as everyone knows, spontaneously emit alpha, beta particles and gamma rays.

The alpha particles emitted by these bodies at prodigious speeds correspond to the nucleus of helium and, therefore, they have a mass 4 and a positive electric charge equal to 2 (considering the mass of hydrogen and the [absolute value of the]\footnote{The words in square brackets have been added (note of the editors).} electrical charge of an electron equal to one), while the beta particles are nothing more than electrons (of negligible mass,\footnote{Here we corrected a wrong statement by the transcribers, who attributed mass 1 to the electron (note of the editors).} which can be approximated to  0, and negative unitary electrical charge)  emitted by radioactive substances with even higher speeds.

The availability of projectiles such as alpha corpuscles  with so large ballistic energy has led the famous English physicist Lord Rutherford for the first time in 1919 to attempt the bombardment of simple materials in the hope that the violent collision produced by an alpha particle with the nucleus of the atom would cause the breakup of the  nucleus itself, varying its characteristics and thus transforming the atom of the studied target into another of a different nature. The eternal dream of the alchemists, the possibility of the transmutation of the elements, became reality with a modern scientific facet.

Lord Rutherford's experiments gave the expected result; this illustrious physicist was able to obtain the disintegration of some elements among those with the lowest atomic number. Nevertheless, experiments carried out on elements of higher atomic weight  did not give the result that could be expected. The explanation is very simple: the alpha particles (in the Rutherford's experiments  nuclei of helium emitted by radium) are positively charged and therefore, when arriving in proximity of the nucleus of the bombarded element, they are electrically repelled due to the positive charge of the  nucleus itself.

Knowing the velocity, mass, and charge of the alpha particle, it is easy to calculate which is the limit element (as a function of its nuclear electrical charge) capable of being disintegrated by the alpha particle.

The experimenters have then turned to different projectiles, for example hydrogen nuclei. The hydrogen nucleus, carrying unitary charge (half the charge of the helium nucleus, that is, of the alpha particle), will undergo half repulsion from the nucleus of the item under study. But the infinitesimal electrical charge of the  hydrogen nucleus is still too big to allow it to get close and collide with the nucleus of an element that, like bismuth (for example), has a nuclear charge 83.

Fortunately, two French experimentalists (Joliot and Curie), bombarding beryllium with alpha particles, got the evidence of a new projectile: the neutron.\footnote{There must be a wrong or incomplete transcription here. According to the transcription, Fermi said, in his Lecture III in Buenos Aires: ``The discovery of the neutron was made in 1932, and is due to a series of works by the German Bothe, the French Curie and Joliot and the English Chadwick.''. Chadwick was credited for the neutron discovery with the Nobel prize in physics in 1935. Joliot and Curie, as Fermi will mention later on, discovered artificial radioactivity, which earned them the Nobel prize in chemistry in 1935 (note of the editors).}
 This corpuscle, emitted by the beryllium atom under the action of the collision of an alpha particle with its nucleus, has a unitary mass and is electrically neutral. 

The advantage of not having any electric charge is evident: the neutron will be able to approach the nucleus of any atom, even those of greater atomic number, without undergoing  electrical repulsion.

On the other hand, the fact of not having an electric charge denies the neutron the possibility to be accelerated by means of electric fields (a method used with alpha and beta particles). To this disadvantage we must also add the  one deriving from the fact that, while the alpha particles are  spontaneously emitted by radioactive bodies, neutrons must be obtained by bombardment and occur in smaller quantity (the number of alpha particles used to obtain the neutrons and the number of neutrons obtained in the experiments carried out in Rome has been $10^{11}$ and $10^{6}$ per second, respectively).

However, this corpuscle has a huge power of penetration, so big that it can go through lead walls more than 30 cm  thick.

After having briefly explained the methods used to obtain the disintegration of the elements, Professor Fermi got directly into the topic of the lecture.

The illustrious speaker has recalled the basis  of his experiments: the phenomenon of artificial radioactivity discovered by the aforementioned physicists Curie and Joliot.

These experimenters, studying the phenomena produced by bombarding  simple bodies with alpha particles, have noticed that in some cases, the disintegration of the nucleus is not immediate: it absorbs the alpha particles to decay spontaneously after a certain time, emitting alpha and beta corpuscles. The analogy between this phenomenon and that of natural radioactivity is evident. For this reason, such phenomenon has been given the name of ``artificial radioactivity".


Let us now see how an atom behaves when an alpha particle collides with its nucleus. Consider the case of  nitrogen: this element has a nucleus whose charge is 7 and whose mass is 14 (we speak here of mass and atomic weight indifferently). Adding to the nitrogen nucleus    an alpha particle (of mass 4 and charge 2) we obtain a total mass of 14 + 4 = 18 and a charge of 7 + 2 = 9. On the other hand, the nitrogen atom, as a result of the collision with the alpha particle, emits a proton
\footnote{Here the transcription says, wrongly, ``an electron''. The error is propagated in the formula. We corrected the mistake (note of the editors).}
(of mass 1 and charge 1), which reduces its mass to 14 + 4 - 1 = 17 and its charge to 7 + 2 - 1 = 8 respectively. Now there is an element whose nuclear electric charge (atomic number) is 8 and whose mass is 17: this well-known element is an isotope of oxygen.

We can then write the nuclear reaction:
\[ \rm{^{14}_{7}N + ^{4}_{2}He} - ^{1}_{1}p = \rm{^{17}_{8}O} \]
where the upper numbers indicate the masses and the lower ones the electrical charges respectively of the nucleus of  nitrogen, the nucleus of helium, the proton and the oxygen nucleus (isotope 17).

By emission of a neutron (of mass one and zero charge) beryllium is transformed into carbon:\footnote{Again a mistake in the transcription, that we corrected (note of the editors).}
\[ \rm{^{9}_{4}Be + ^{4}_{2}He} - ^{1}_{0}n =\rm{^{12}_{6}C} \, . \]

Having clarified these concepts, Professor Fermi told us that, in addition to aluminium, boron and magnesium, also other elements could undergo the phenomenon of artificial radioactivity. 

He has started in Rome in March of this year a series of investigations carrying out a thorough attack on the elements even among the heaviest ones using neutrons as bombardment projectiles.

Neutrons have been obtained with the following method.
A certain amount of pulverized beryllium and a quantity of radium emanation corresponding to approximately 800 milliCurie were closed in a tube
through condensation with the help of liquid air. Once closed, the tube is ready as a neutron generator. The alpha particles emitted in great quantity by the radioactivity (approximately $10^{11}$ per second, as  said) colliding with the nuclei of the beryllium atoms originate the neutrons, which  serve to bombard the substance under study. For this purpose, the substance is arranged around the  tube, letting it stay for a time long enough  to obtain the desired phenomenon in an appreciable magnitude.

Exposing a body to neutron bombardment can lead to three different results:
\begin{enumerate}
\item The element is inert to the bombardment and there is no transformation.
\item The element is transformed into another of greater or lesser atomic weight, but perfectly stable (this is the case of the transformation of the nitrogen into oxygen). In this case the phenomenon of artificial radioactivity does not take place and, since the quantities of transformed matter are infinitesimal, the experimental method does not allow us to see any effect.
\item The element is transformed into another that is unstable and, therefore, gives rise to the phenomenon of artificial radioactivity.
\end{enumerate}

The last one is the most interesting case and we will deal with it in the following.

To study this phenomenon, it will initially be necessary to have an apparatus capable of revealing the phenomenon itself. This is normally obtained (and so did Professor Fermi) using the Geiger and M\"uller counter, i.e., an ingenious discharge chamber, whose sensitivity is so good that it allows to detect quantities  as small as a ``quantum" of light.

In Professor Fermi's experiments, the counter consisted of a small aluminium tube whose ends are closed by means of insulating material caps  that hold a thin aluminium wire along the axis of the tube. The walls of this counter  are very thin (from one to two tenths of a millimeter) such as, however, to prevent the entry of alpha particles  but to allow the passage of the beta particles (which are more penetrating).

The pressure inside the tube is reduced to values corresponding to a few centimeters of mercury. The central electrode (the aluminium wire) is set to zero potential while the wall is brought to potentials very close to that which would be able to induce the electric discharge. In Professor Fermi's experiments this potential, given the dimensions of the counter and its internal pressure, ranged from 1000 V to 2000 V.

Let us see now what is the phenomenon that occurs when a radioactive substance, that is, a substance capable of spontaneously emitting  alpha and beta particles, approaches the counter. The beta particles, due to their enormous speed, pass through the walls of the counter, penetrating it and producing the well-known phenomenon of ionization along their path. This ionization obviously favors the electric discharge that takes place inside the counter during the infinitesimal amount of time that the beta particle spends to cross it. By connecting the counter to a loudspeaker through a suitable amplifier, the experimenter will be able to hear the sound of a discharge every time a beta particle reaches the inside of the tube. It is now easy to record the number of discharges that take place every second and to indirectly know the greater or lesser activity of the body under study and the speed of its evolution period. It is evident that the number of discharges will be influenced by all causes capable of producing ionization (X-rays, cosmic radiation, traces of radioactive substances contained in the material with which the counter is built, etc.), but a previous calibration of the instrument allows knowing with sufficient accuracy the part that in the mentioned phenomenon corresponds to the artificial radioactivity of the body under study.


This body, exposed to neutron bombardment for a reasonable time, is successively brought in close proximity to the Geiger counter, which reveals whether the phenomenon of transformation of the initially radioactive element has actually taken place. When this phenomenon is observed and once the evolution period has been determined, it will be necessary to proceed to a chemical investigation capable of identifying the nature of the new element generated. This investigation is one of the most delicate because the order of magnitude of the quantities involved is so small (it corresponds to a very limited number of atoms) that the normal procedures of analytical chemistry are not sufficient to precipitate the desired element. An ingenious trick is then used: the body produced must be an isotope of one of the elements next to the element under study (in  Mendeleev table) or an unknown new element. In the first case (radioactive isotope), due to the chemical affinity of the isotopes, before proceeding to the precipitation of the alleged element, small amounts of salts of the elements are added to the solution under study, among which, with greater probability, the isotope of the element generated in the bombardment would be included. An example will serve to make things clearer: Let us suppose that the element under study is chlorine, which, under the action of neutron bombardment, gives rise to an element with radioactive properties. Chlorine ranks as follows in  
Mendeleev table:\footnote{Here the transcription says, wrongly, Ag, silver, instead of Ar, argon (note of the editors).}
\[ \rm{P; \; S; \;  Cl;\;  Ar; \; K} \]
that is, it is preceded by sulfur and phosphorus and followed by argon and potassium.

It is highly probable, therefore, that the element generated by chlorine is an isotope of one of those that surround it and, in fact, by taking a phosphorus salt in solution and precipitating this element again, a substance is obtained (based on phosphorus isotopes), which, brought in proximity to the Geiger counter, shows to possess the same characteristics (activity and evolution period) as the activated chlorine. This fact allows one to conclude that chlorine, under the action of neutron bombardment, is transformed into a new radioactive element, corresponding to an isotope of phosphorus. The nuclear reaction confirms this result:
\[ \rm{^{34}_{17}Cl}  + ^{1}_{0}n - \rm{^{4}_{2}He}   = \rm{^{31}_{15}P} \, . \]

In some cases, the phenomenon is complicated by giving rise to two or more radioactive elements at the same time, which can be separated only due to the difference in the corresponding decay periods.

The phenomena that we have just summarized have made Professor Fermi think about the possibility of increasing even more the already intense activity of some spontaneously radioactive substances by means of neutron bombardment.

Thus we come to the case of uranium, element 92 of  Mendeleev table, which through successive emissions of alpha and beta particles is transformed into uranium U 
X$_1$ (ejection of an alpha particle), in uranium U X$_2$ (ejection of a beta particle), in uranium U$_{\rm{II}}$ (ejection of a beta particle), 
etc., until reaching radium and, finally, polonium [and then lead]:\footnote{The $^{238}$U atom was historically called U$_{\rm{I}}$. The formula, which was not explicitly present in the original transcription,  shows in parentheses the historical names used for the isotopes. The words in square brackets have been added. Here it has been very difficult for the transcribers to follow correctly (note of the editors).}  
\[ \rm{^{238}U \xrightarrow{\alpha} \, ^{234}Th \, (U\, X_1)
\xrightarrow{\beta} \, ^{234}Pa \, (U\, X_2)
\xrightarrow{\beta} \, ^{234}U \, (U_{II})
\xrightarrow{} (...) \, \xrightarrow{} \, ^{210}Po \, .
}\]

Experimentation on this element presents enormous difficulties, especially due to the fact that the natural transformation of uranium into uranium X$_1$ (the one that emits beta particles) has a relevant effect on the Geiger counter, in the sense of overlapping the signal due to the natural radioactive phenomenon to that of the artificial phenomenon.

However, the difficulties have been overcome by a scrupulous purification of the uranium (elimination of U X$_1$ and U X$_2$) and by carrying out the experiment as quickly as possible.

The results of the neutron bombardment have been surprising: the Geiger counter has revealed four new radioactive phenomena unknown until then, phenomena that reveal the existence of new radioactive products whose evolution period (the time necessary to reduce by half the value of the product's own radioactivity) is (approximately) 4 seconds, 10 seconds, 13 minutes and 100 minutes respectively. No analysis could be carried out on the elements with a period of 4 and 10 seconds, because the radioactivity of these elements practically canceled before  the chemical separation could be achieved (although these separations are carried out in a ``record" time, they always require a minimum of 2 to 3 minutes). Therefore, nothing can be claimed about them to date.

The element with a period of 100 minutes is in almost the same situation because, during this period of time, the uranium is partially transformed into uranium U X$_1$ which, emitting beta particles,  can in turn undergo  the  phenomena which are  possible for the new element.

The most interesting case is that of the 13-minute period element, allowing all necessary analyses to be carried out comfortably. The method used for its identification is   that already described for chlorine (adding and precipitating salts of probable isotopes), and here something happens on which all the theories and opinions expressed on the possible discovery of 
the new element 93 are based.

Uranium is element 92 in   Mendeleev table and, therefore, has the highest atomic weight among those known. 

The elements that precede it (and among which  the corresponding isotope should be searched for) are indicated in the following list where each element is preceded by its  atomic number:
\[ \rm{84 \ Po - 85 \ ? - 86\ Em - 87\ ? - 88 \ Ra - 89\ Ac - 90 \ Th - 91 \ Pa - 92\ U} \, . \]
Chemical analysis has been attempted by successively adding salts of Pa, Th, Ac, Ra, Em and Po without precipitating the hitherto unknown element.

At this point the illustrious speaker expresses major reservations. Before being able to affirm that we are in front of  a new element, and precisely element 93, Professor Fermi states that it is necessary to be sure that the element in question does not correspond to an isotope of any of the 91 elements that precede uranium. This fact is very unlikely (according to the eminent experimenter's opinion) especially since it has been verified that the characteristics of this element do not correspond to either the unknown element 87 (which should be gaseous) or element 85 (which should  necessarily present the features of a halogen), but we are not yet authorized to affirm that the element in question corresponds to 93 or perhaps 94 or 95 on the atomic scale. Ongoing experiments to obtain a complete chemical picture of this mysterious element will tell us the final word on such an exciting question.

\newpage

\section*{Radiation\\(La Plata, August 17, 1934)}

\vskip 0.3 cm

\addcontentsline{toc}{section}{Enrico Fermi's lecture in La Plata (August 17)}

\addcontentsline{toc}{subsection}{Radiation (radioactivity, light  and cosmic rays)}

On August 17, 1934, a  special reception for  Professor Enrico Fermi,  academician of Italy, was held at the National Academy of Agronomy  
and Veterinary
  in the  Wenceslao Escalante lecture hall of the  corresponding Faculty,  attended by the Dean, Doctor Zanolli,  by the President of the Instituto
Argentino de Cultura It\'alica,  Doctor  Marotta,
 and by numerous
academicians, teachers and students.

Professor Fermi lectured on ``Radiation". 
The President of the Academy, Engineer Marotta, welcomed him and, after presenting
  the Argentinean Institute of Italian Culture,  highlighted the figure of Doctor Fermi, reviewing his research and discoveries. 
He said that his lecture was  useful not only for the impact of  his experimental studies on the disciplines of 
agronomists and veterinarians, but also because the University, apart from
its institutional and research role,  should also take care
of the general culture of the graduates. He recalled  in this regard the position of
Ortega y Gasset on the need for everyone to possess, whatever
the specialization of his studies, an idea of the physical, biological,
historical and philosophical conception of the world, highlighting in this regard the importance
of physics.

Professor Fermi gave an outstanding lecture from which
it has only been possible to extrapolate the brief summary that follows.

After reminding the ever-increasing importance of the study of
radiation, not only in the domain of physics but also of 
biology, Professor Fermi stated that for reasons of
competence he would focus on the physical aspect of the problem. 
He discussed
 briefly  the development of studies on luminous radiation,  that gradually led to the knowledge of other
types of radiation, not directly perceptible to our eyes, but
that have a  nature analogous to that of light. He then discussed the origin
of the remarkable differences that exist in the physical and chemical effects of all these types of rays, 
proving the theory that
  the frequency of the radiation is the key element
for the production of various reactions.

He then reviewed other types of radiation consisting of
very fast corpuscles.
These radiations can be classified according to the corpuscles
(electrons, alpha particles, neutrons, etc.) they are made of  and according to their
velocity. Doctor Fermi  explained how, based on these
 elements, it is possible to understand [their action on]\footnote{The words in square brackets are missing in the original text in Spanish, certainly by mistake (note of the editors).} the substances they reach, this
 action consisting mainly in ionizing the
atoms; he argued that, ultimately,  most of the biological actions of
the various types of radiation must surely be linked
to side effects of ionization.

Finally, he addressed the problem of cosmic radiation, examining
 the main hypotheses that have been proposed to
explain it. It is --he said-- a radiation that, although  of very weak intensity, thanks to the enormous energy of the corpuscles that
 constitute it can sometimes produce disintegration phenomena
of much greater magnitude than can be obtained in the laboratory.


\begin{thebibliography}{}
\newcommand{\BY}[1]{{#1}}
\addcontentsline{toc}{section}{Bibliography}
\bibitem{treccani} The biographical information is taken from the site of the Enciclopedia Italiana http://www.treccani.it/.
\bibitem{bernardini} \name{C. Bernardini and L. Bonolis (eds.)} \book{Enrico Fermi: His Work and Legacy
(Societ\`a Italiana di Fisica, Bologna, and Springer-Verlag, Berlin, 2004).}
\bibitem{betars} \name{E. Fermi} {\em Tentativo di una teoria dei raggi $\beta$,} \rev{La Ricerca Scientifica}{2}{12}{1933}
\bibitem{betanc} \name{E. Fermi} {\em Tentativo di una teoria dei raggi $\beta$,} \rev{Nuovo Cimento}{11}{1}{1934}
\bibitem{ministero} Centro de Estudios Migratorios Latinoamericano (CEMLA), Base de datos de inmigrantes.
\bibitem{corriere} Corriere della Sera, July 24 1934, page 3.
\bibitem{laura} \name{L. Fermi} \book{Atoms in the Family: My Life with Enrico Fermi (University of Chicago Press, 1958).}
\bibitem{critica} Cr\'itica, July 30 1934, page 1.
\bibitem{mattino} {\em La prima conferenza del Prof. Fermi alla Facolt\`a Universitaria di Scienze,} Il Mattino d'Italia, August 3 1934.
\bibitem{segre} \name{E. Segr\`e} \book{Enrico Fermi, physicist (University of Chicago Press, 1970).}
\bibitem{ba} Conferencias, Facultad de Ciencias exactas F'sicas y Naturales, Publicaci\'on 15, Buenos Aires, 1934.
\bibitem{opere} \name{E. Amaldi, H.L. Anderson, E. Persico, F. Rasetti, C.S. Smith, A. Wattenberg, E. Segr\`e (eds.)} \book{Enrico Fermi collected works (University of Chicago Press and Accademia Nazionale dei Lincei, Rome, 1962).}
\bibitem{cordoba} \name{E. Fermi} {\em Experimentos de radioactividad artificial,} \rev{Revista de la Universidad Nacional de C\'ordoba}{5/6 (21)}{187}{1934}
\bibitem{laplata} \name{E. Fermi} {\em Radiaciones,} \rev{Anales de la Academia Nacional de Agronom\'ia y Veterinaria, Universidad Nacional de La Plata}{Tomo I 1932--1934}{405}{1934}
\bibitem{brazilians} \name{F. Caruso and A.J. Marques} {\em Sobre a viagem de Enrico Fermi ao Brasil em 1934,} \rev{Estudos Avan\c{c}ados}{28}{279}{2014}
\bibitem{brazilians2} \name{F. Caruso and A.J. Marques}  {\em A visita de Fermi ao Brasil segundo os jornais brasileiros da \'epoca,} \rev{Scientiarum Historia}{VI}{}{2013}
\bibitem{intropart} \name{A. De Angelis and M. Pimenta} \book{Introduction to particle and astroparticle physics, 2$^{nd}$ ed. (Springer-Nature, Heidelberg, 2018).}

\end{thebibliography}
\end{document}